\documentstyle[12pt,graphicx,caption2]{article}

\begin{document}

\title{Field and phenomenological description of absorption of particles}
\author{V.I.Nazaruk\\
Institute for Nuclear Research of RAS, 60th October\\
Anniversary Prospect 7a, 117312 Moscow, Russia.\\
 e-mail:nazaruk@inr.ru}

\date{}
\maketitle
\bigskip

\begin{abstract}
Complicated processes in the medium (decays, reactions and $n\bar{n}$ 
conversion) involving final state absorption are considered. The 
calculations in the framework of field and phenomenological approaches 
are compared. The reasons for disagreement are studied. The field 
approach can tend to increase of total process probability as well as 
probability of channel corresponding to absorption, in comparison with 
phenomenological model. 
\end{abstract}

\vspace{5mm}
 
\newpage
\setcounter{equation}{0} 
The multistep processes in the medium are frequently attended by 
the absorption of particles in the intermidiate or final states. 
The simplest two-step process of this type is the $n\bar{n}$ 
conversion [1,2] in nuclear matter: 
\begin{equation}
n\rightarrow \bar{n}\rightarrow M,
\end{equation}
here $M$ are the annihilation mesons. The qualitative process picture is as
follows. The free-space $n\bar{n}$ conversion comes from the exchange of Higgs
bosons with a mass $m_H>10^5$GeV [1,2]. From the dynamical point of view this
is a momentary process: $\tau \sim 1/m_H<10^{-29}$ s. The antineutron
annihilates in a time $\tau \sim 1/\Gamma \sim 10^{-24} $s, where $\Gamma $
is the annihilation width of $\bar{n}$; $\Gamma \sim 100$ MeV.

Phenomenologically, the absorption of particles in the medium is described by
$ImU_{opt}$, where $U_{opt}$ is the optical potential. First of all we consider
the antineutron annihilation by means of field approach. Let $U_n$ and
$U_{\bar{n}}$ are the real potentials of $n$ and $\bar{n}$, respectively.
The background nuclear matter field $U_n$ is included in the neutron wave
function: $n(x)=V^{-1/2}\exp (-i({\bf p}_n^2/2m+U_n)t+i{\bf p}_n
{\bf x})$. The interaction Hamiltonian is
\begin{equation}
{\cal H}_I={\cal H}_{n\bar{n}}+V\bar{\Psi }_{\bar{n}}\Psi_{\bar{n}}+{\cal H}_a,
\end{equation}
\begin{equation}
{\cal H}_{n\bar{n}}=\epsilon \bar{\Psi }_{\bar{n}}\Psi _n+H.c.,
\end{equation}
$V=U_{\bar{n}}-U_n$. Here ${\cal H}_{n\bar{n}}$ and ${\cal H}_a$ are the
Hamiltonians of $n\bar{n}$ conversion and annihilation, respectively; $\epsilon $
is a small parameter.

The corresponding diagram is shown in Fig. 1. The antineutron Green's function
has the form
\begin{equation}
G=\frac{1}{G_0^{-1}-V}=-\frac{1}{V},
\end{equation}
$G_0^{-1}=\epsilon _n-({\bf p}_n^2/2m+U_n)=0$. The process amplitude is
\begin{equation}
M=-\epsilon GM_a=\epsilon \frac{1}{V}M_a.
\end{equation}
Here $M_a$ is the annihilation amplitude. For the process width $\Gamma _f$
one obtains
\begin{equation}
\Gamma _f=N\int d\Phi \mid\!M\!\mid ^2=\frac{\epsilon ^2}{V^2}N\int d\Phi
\mid\!M_a\!\mid ^2=\frac{\epsilon ^2}{V^2}\Gamma .
\end{equation}
The normalizing multiplier $N$ is the same for $\Gamma _f$ and $\Gamma $.

\begin{figure}[h]
  {\includegraphics[height=.25\textheight]{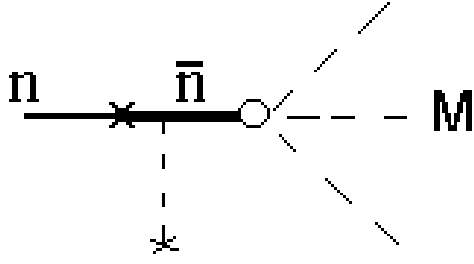}}
  \caption{$n\bar{n}$ transition in nuclear matter. The antineutron
annihilation is illustrated by circle.}
\end{figure}

In the phenomenological approach the $\bar{n}$-medium interaction is
described by optical potential $U_{opt}$. Instead of Hamiltonian (2) we
have
\begin{equation}
{\cal H}_I={\cal H}_{n\bar{n}}+V\bar{\Psi }_{\bar{n}}\Psi_{\bar{n}}-
i\Gamma /2\bar{\Psi }_{\bar{n}}\Psi_{\bar{n}},
\end{equation}
$ImU_{opt}=-\Gamma /2$, $ReU_{opt}=V$. As a rule, this model is realized by
means of equation of motion [3-5]. To avoid questions connected with the
formalism, we use the diagram technique, as for the model (2). Certainly,
in the both cases the results coincide.

The total process width $\Gamma _p$ is obtained from the unitarity condition:
\begin{equation}
\Gamma _p=\frac{1}{T_0}(1-\mid S_{ii}\mid ^2)\approx \frac{1}{T_0}2ImT_{ii},
\end{equation}
$S=1+iT$. Here $T_0$ is the normalizing time, $T_0\rightarrow \infty $. (In
the general case $\Gamma _p=\Gamma _{\bar{n}}+\Gamma _a$, where $\Gamma _
{\bar{n}}$ and $\Gamma _a$ correspond to antineutron and annihilation mesons
in the final state, respectively. For the $n\bar{n}$ conversion in nuclear
matter $\Gamma _{\bar{n}}\approx 0$.)

In the impulse representation
\begin{equation}
\Gamma _p=-2Im\epsilon G\epsilon =2Im\epsilon \frac{1}{V-i\Gamma /2}\epsilon.
\end{equation}
Finally
\begin{equation}
\Gamma _p=\epsilon ^2\frac{\Gamma }{V^2+(\Gamma /2)^2}\approx 4\epsilon ^2
/\Gamma .
\end{equation}

Since all the antineutrons annihilate, we can compare $\Gamma _p$ and $\Gamma _f$:
\begin{equation}
\frac{\Gamma _f}{\Gamma _p}=1+\left(\frac{\Gamma /2}{V}\right)^2>1.
\end{equation}

The field approach gives reinforcement, in comparison with the phenomenological
model. Moreover, the $\Gamma $-dependence of Eqs.(10) and (6) is inverse:
\begin{equation}
\frac{d\Gamma _f}{d\Gamma }>0.
\end{equation}
We stress that for any model of $\bar{n}$-medium interaction field approach
gives $\Gamma _f\sim \Gamma $. 

If $(\Gamma /2)^2>V^2$ (the realistic set of parameters [4] fits this
requirement),
\begin{equation}
\frac{d\Gamma _p}{d\Gamma }<0.
\end{equation}
At the point $(V=0,\Gamma )$ Eq.(13) is true as well. Therefore, the 
phenomenological model gives the inverse $\Gamma $-dependence. At the same 
time annihilation is the main effect determining the process speed (see 
Eqs.(6) and (10)).

The phenomenological model is based on Eqs.(7) and (8). If instead of
Hamiltonian (7) we take
\begin{equation}
{\cal H}_I={\cal H}_{r,d}+{\cal H},
\end{equation}
\begin{equation}
{\cal H}=-i\Gamma /2\bar{\Psi }_{\bar{n}}\Psi_{\bar{n}},
\end{equation}
where ${\cal H}_r$ and ${\cal H}_d$ are the Hamiltonians of reaction and
decay with $\bar{n}$ in the final state, respectively, the qualitative 
conclusions do not change because the heart of the problem is in the term 
${\cal H}$. In the next paper these problems will be studied in detail.

From Eq.(10) it is seen that real potential $V$ leads to suppression, which is
certainly correct. However, $\Gamma $ acts in the same direction, which seems
wrong at least at the threshold point $\Gamma =0$ because the procedure $\Gamma
=0 \rightarrow \Gamma \neq 0$ implies the opening of a new channel of
$\bar{n}$-medium interaction (annihilation). This must tend to increase
$\Gamma _p$. Eq.(13) provides inverse tendency.

The reason for disagreement indicated above is the incorrect description
of absorption on the whole, in particular the nonhermicity of Hamiltonian (7).
Eq.(8) follows from the unitarity condition. Since the operator $iImU_{opt}$ is
"antihermitian", we have
\begin{equation}
(SS^+)_{fi}=\delta _{fi}+\alpha _{fi},
\end{equation}
$\alpha _{fi}\neq 0$. Then instead of equation
\begin{equation}
\sum_{f\neq i}\mid T_{fi}\mid ^2\approx 2ImT_{ii}
\end{equation}
one obtains
\begin{equation}
\sum_{f\neq i}\mid T_{fi}\mid ^2=2ImT_{ii}-\mid \!T_{ii}\!\mid ^2+\alpha
_{ii}\approx 2ImT_{ii}+\alpha_{ii},
\end{equation}
$\alpha _{ii}\neq 0$. Let us consider the right-hand side of this equation.

To get the expression for observable values we pass on to the evolution
operator $U(t)=1+iT(t)$. Then
\begin{equation}
\sum_{f\neq i}\mid T_{fi}(t)\mid ^2\approx 2ImT_{ii}(t)+\alpha_{ii}=W(t),
\end{equation}
where $W(t)$ is the process probability in a time t. $\alpha _{ii}$ can
be neglected only when $\mid \!\alpha _{ii}\!\mid \ll 2ImT_{ii}$. However,
for the value of $2ImT_{ii}(t)$ we have obtained (see Eqs.(8)-(10))
\begin{equation}
W(t)=\Gamma _pt=\frac{4\epsilon ^2}{\Gamma }t.
\end{equation}
We consider the $n\bar{n}$ transition in nuclear matter. We take $\Gamma =
100$ MeV, $t=T_0=1.3$ yr [6] ($T_0$ is the observation time in proton-decay
type experiment) and $\epsilon =1/\tau <1/(10^8 s)$ [7,8]. As a result
$2ImT_{ii}<10^{-31}$. We believe that $10^{-31}\ll\mid \!\alpha _{ii}\!\mid $
(in our opinion $\alpha \sim 1$) and the basic equation $W(t)=2ImT_{ii}(t)$
and thus Eq.(8) are meaningless.

Eq.(16) implies that: (a) Any matrix element $S_{nm}$ (on-diagonal and
off-diagonal) can contain some error. (b) The basic relation (8) is broken
down. So the physical meaning of imaginary part of self-energy $Im\Sigma =
-\Gamma /2$ (see Eq.(9)) is uncertain because it is clarified using the
relation (8).

Strictly speaking, Eq.(8) can be used only for unitary $S$-matrix, or
equations of Schrodinger type (when the interaction Hamiltonian contains the
single term ${\cal H}_I={\cal H}$) as in this case the problem is unitarized:
by means of equation of motion and condition of probability conservation $1=
\mid U_{ii}(t)\mid ^2+W(t)$ the optical potential is fitted to experimental
data. Certainly, the unitarity is a necessary and not a sufficient condition.

The particle absorption in the medium can be considered as a decay of
one-particle state. With substitution
\begin{equation}
iImU_{opt}=-i\Gamma _x /2,
\end{equation}
where $\Gamma _x$ is a width of some free-space decay $\bar{n}\rightarrow x$,
the formulas given above describe the decay in a final state (instead of
absorption). Formally, in this case all the results are true as well.

Also we would like to stress the following. The importance of unitarity
condition is well known [9,10]. Nevertheless, the nonhermitian models (7),(8)
and (14),(8) are frequently used since they strongly simplify the calculations. 
In particular, all existing calculations of process (1) are based on model
(7),(8) (see [5] for future references), resulting in the inverse
$\Gamma $-dependence.

In the general case the total process probability corresponding to Eq.(14) is 
\begin{equation}
W_t=W_{\bar{n}}+W_a,
\end{equation}
where $W_{\bar{n}}$ and $W_a$ are the probabilities to find antineutron and
annihilation products, respectively. The phenomenological Hamiltonians like
(14) containing several terms describe only $W_{\bar{n}}$. This result as
well as other arguments in favour of tendency $d\Gamma _f/d\Gamma >0$ will
be presented in the next paper. The range of applicability of the
complicated models containing the Hamiltonian ${\cal H}$ will be studied
as well.

The main result is as follows: the absorption (decay) in a final state
does not tend to the process suppression. This is true for the reactions,
decays and $n\bar{n}$ conversion in the medium. The field approach gives
$\Gamma _f\sim \Gamma $, whereas models (7) and (8) give $\Gamma _p\sim
1/\Gamma $. Therefore, the calculations based on field approach (unitary
models) can give a reinforcement of the corresponding processes (see
Eq.(11)), in comparison with phenomenological model results.\\
\\
1. S.L. Glashow, Preprint HUTP-79/A059 (Harvard, 1979).\\
2. R.E. Marshak and R.N. Mohapatra, Phys. Rev. Lett. {\bf 44}, 1316 (1980).\\
3. P.G.H. Sandars, J.Phys. {\bf G6}, L161 (1980).\\
4. C.B. Dover, A.Gal and J.M.Richard, Phys. Rev. {\bf C31}, 1423 (1985).\\
5. V.I. Nazaruk, Phys. Lett. {\bf B337}, 328 (1994); Phys. Rev. {\bf C58},

R1884 (1998).\\
6. H. Takita et al.(Kamiokande), Phys. Rev.{\bf D34}, 902 (1986).\\
7. V.I. Nazaruk, Yad. Fiz. {\bf 56}, 153 (1993).\\
8. M. Baldo-Ceolin et al., Phys.Lett. {\bf B 236}, 95 (1990).\\
9. T.Inone and E.Oset, Nucl. Phys. {\bf A 721}, 661 (2003).\\
10.E.Oset, Nucl. Phys. {\bf A 721}, 58 (2003).

\end{document}